# Anisotropic Deformation in the Compressions of Single Crystalline Copper Nanoparticles

Jianjun Bian[1,3], Hao Zhang[2], Xinrui Niu[3] and Gangfeng Wang[1,*]

[1]    Department of Engineering Mechanics, SVL, Xi'an Jiaotong University, Xi'an 710049, China

[2]    Department of Chemical and Materials Engineering, University of Alberta, Edmonton, T6G 1H9, Canada

[3]    CASM and Department of Mechanical and Biomedical Engineering, City University of Hong Kong, Hong Kong SAR, China

[*]    Correspondence: wanggf@mail.xjtu.edu.cn

## Abstract

Atomistic simulations are performed to probe the anisotropic deformation in the compressions of face-centred-cubic metallic nanoparticles. In the elastic regime, the compressive load-depth behaviors can be characterized by the classical Hertzian model or flat punch model, depending on the surface configuration beneath indenter. On the onset of plasticity, atomic-scale surface steps serve as the source of heterogeneous dislocation in nanoparticle, which is distinct from indenting bulk materials. Under [111] compression, the gliding of jogged dislocation takes over the dominant plastic deformation. The plasticity is governed by nucleation and exhaustion of extended dislocation ribbons in [110] compression. Twin boundary migration mainly sustain the plastic deformation under [112] compression. This study is helpful to extract the mechanical properties of metallic nanoparticles and understand their anisotropic deformation behaviors.

# 1. Introduction

Single crystal nanoparticles play increasing important roles in a wide variety of fields such as fuel cells, energetic materials and high performance catalyst. Among these applications, some crucial functions are closely in coordination with the mechanical properties of nanoparticles. It is well known that the mechanical properties of such surface-confined materials are drastically different from their macro-scale counterparts. For example, both the Young's modulus and hardness of single crystal nanoparticles are a bit larger than those of their bulk counterparts [1,2]. Extensive efforts have been devoted to investigate the unique properties of nanoparticles.

Anisotropy is one inherent characteristic of crystalline materials, and strongly affects their elastic and plastic deformation. For bulk single crystal materials, the reduced Young's modulus, the location of nucleation sites and the nucleation stress vary with different lattice orientations under indentation [3]. In incipient plasticity, dislocation nucleation depends largely on the available slip systems [5]. Anisotropic effects become even more prominent in low dimensional nanostructures such as nanowires and nanopillars. For example, lattice orientation in axial direction significantly affects the yield stress of gold nanowires [6]. The exhibiting tension-compression asymmetry in nanopillars depends on crystallographic orientation [7]. The plastic deformation of nanowires under torsion can be either homogeneous or heterogeneous, depending on the wire orientation [8]. In the case of silicon nanowires under uniaxial tension, the fracture mechanism would switch from brittle to ductile for varying axial crystallographic orientation [9]. The prominent anisotropy in low dimension materials is closely related to the intrinsic crystal structure and the extrinsic surface morphology. Free surfaces usually serve as preferential dislocation nucleation sites. Changing surface structure by removing weakly bound atoms produces a striking rise in yield strength [10].

For nanoparticles, such factors as surface facets, geometric profile and internal twin boundaries contribute to their unique properties [11]. Extensive experiments and simulations were performed to investigate the deformation mechanisms in nanoparticles. For example, phase transition was observed in single-crystal silicon nanoparticles during uniaxial compression [13]. The *in situ* TEM indentation of silver nanoparticle revealed reversible dislocation plasticity in nanoparticles [14]. During uniaxial compression, lateral free surface strengthened defect-free gold nanoparticles by draining dislocations from particles [15]. For uniaxial [001] compression of spherical copper nanoparticles, deformation twinning dominated the severe plastic deformation [16]. Microstructural evolution of tin dioxide nanoparticles under compression exhibited the formation of shear bands, twinning and stacking faults [17].

Up to now, there is still a lack of systematic investigation of anisotropic behaviors of face-centred-cubic (fcc) metallic nanoparticles. In the present study, we conduct molecular dynamics (MD) simulations to illuminate both the elastic characterization and plastic deformation mechanisms of copper nanoparticles, aiming to present a landscape of anisotropic deformation of low dimensional nanostructures.

## 2. Methods

MD simulations are conducted using the open-source simulator LAMMPS developed by Sandia National Laboratories. The embedded atom method (EAM) model is utilized to describe the atomic interaction of copper atoms. According to the EAM model, the total energy $U$ of a system containing $N$ atoms is expressed as

$$U = \sum_{i=1}^{N} \left( F_i(\bar{\rho}_i) + \frac{1}{2} \sum_{i \neq j}^{N} \phi_{ij}(r_{ij}) \right),$$ (1)

where $F_i(\bar{\rho}_i)$ is the embedding energy depending on atomic electron density $\bar{\rho}_i$ at the position of the $i$-th atom, $\phi_{ij}(r_{ij})$ is the pair-wise interaction energy related to the interatomic distance $r_{ij}$ between atoms pairs $i$ and $j$. In the present study, we use an EAM potential for copper parameterized by Mishin *et al* [18], which was constructed by fitting both experimental and *ab-initio* computational data, for instance, cohesive energy, bulk modulus, elastic constants, intrinsic stacking fault energy, and vacancy formation and migration energies. This potential has been widely used to investigate mechanical properties and deformation mechanisms in different nanostructures [19].

The simulation model of uniaxially compressing a nanoparticle is depicted in Figure 1. One single crystal copper nanoparticle with radius of 10 nm is carved out of bulk defect-free single crystal copper, and contains about ~ 0.35 million atoms. Then the particle is placed between two rigid planar indenters. A repulsive potential is utilized to model the frictionless compression as

$$U_i\left(z_i\right)=\begin{cases} K\left(z_i-h_T\right)^3 & z_i \geq h_T \\ 0 & h_b < z_i < h_T \\ K\left(h_B-z_i\right)^3 & z_i \leq h_B \end{cases}, \qquad (2)$$

where $K$ is a specified constant representing the rigidity of the planar indenter. Compression direction is parallel to the $z$-axis, and $z_i$, $h_T$ and $h_B$ represent the positions of the $i$-th atom, the top indenter, and the bottom indenter, respectively. Based on the previous studies [16], we chose three typical orientations with lower Miller indices, [110], [111] and [112], as the uniaxial compression directions in the present study.

Loading procedure is implemented within the framework of canonical (NVT) ensembles. Before compression, the as-carved spherical nanoparticle firstly is performed structure relaxation using the conjugate gradient method, and then is equilibrated at 10 K for about 20 ps to relief the internal stress. When compression is conducted, the top and bottom planar indenters are simultaneously move towards the center of nanoparticle with a speed of ~ 0.1 Å/ps, and the compression depth $\delta$ is denoted by the displacement of one indenter. During the loading process, the temperature is controlled at 10 K using a Nosé-Hoover thermostat, and the time step of velocity-Verlet integration is chosen as 2.0 fs.

To identify the characteristics of nucleated defects inside the particles and visualize their evolution processes, the local crystal structure of each atom is computed based on the common neighbor analysis (CNA) [21]. Atomic configurations and defect structures are visualized using Ovito [22] and Paraview [23]. To reveal the mechanical response of nanoparticles, the variations of contact area and averaged contact stress are examined. Contact area is determined by the Delaunay triangulation algorithm, and the averaged contact stress is defined as the loading force divided by contact area.

## 3. Results and Discussions

Compression of nanoparticle clearly demonstrates that the elastic response, initial dislocation nucleation and the following defects evolution vary with different loading directions. These orientation dependent features originate from the intrinsic crystallographic structure, surface configurations and the activated slip systems. In this section, we will analyze these behaviors in details.

### 3.1 [111] Compression

Figure 2(a) shows the loading response under [111] compression. The overall load response can be clearly subdivided into elastic and plastic stages, connected by the critical yielding point. In the elastic stage, compression load linearly increases with the accumulation of compression depth. After reaching the yielding point, the load is abruptly punctuated by an evident drop, indicating the initiation of plastic deformation. In the plastic stage, load fluctuates serratedly with further compression, as a reflection of inside atomic defect evolutions. Figure 2(b) demonstrates the variations of the averaged contact stress and contact area with respect to compression depth. In the elastic stage, the contact stress obtains its maximum value as high as ∼ 26.0 GPa, followed by a sudden drop with the occurrence of yielding. Under further compression, contact stress fluctuates and maintains at a low level. Contact area is constant in elastic stage, and exhibits a stepwise increasing at yielding point, and then keeps increasing over the rest loading stage.

Atomic surface structures govern the elastic response of nanoparticles. Along [111] direction, atomic-scale facets are the most prominent characteristics of spherical surface. When compression begins, planar indenters firstly touch the outmost facet, rather than an ideal smooth spherical surface. It is manifested in recent studies that the flat punch model should be utilized when surface step is compressed [24,25]. In this model, compressive load $F$ is expressed as a function of compression depth $\delta$ by

$$F = 2E^* a\delta,\qquad(3)$$

where $a$ and $E^*$ are the contact radius and the reduced modulus, respectively. In Figure 2(a), we use Eq. (3) to fit the loading curve in the elastic stage, and the fitted reduced modulus is ∼ 201.2 GPa. In the [111] nanoindentation of bulk copper, the reduced modulus is extracted as ∼193 GPa [5]. It is noted that nanoparticle have a slightly larger elastic modulus than that of bulk material, which may be attributed to surface effects. Based on this model, the theoretical predictions of contact stress and contact area are plotted in Figure 2(b), which coincide well with the MD computational results in elastic stage.

From yielding point, dislocation initiates inside the nanoparticle. The slip systems available for [111] compression is depicted in Figure 3(a), and initial surface morphology and surface steps are displayed in Figure 3(b). The outmost surface facet has a hexagonal shape, whose edges are preferential sites for dislocation nucleation. When yielding occurs, heterogeneous dislocations nucleate around surface steps as depicted in Figure 3(c). Embryo size is indicative of the sequence of dislocation nucleation, and three larger embryos as one set (marked by blue arrows) nucleate earlier than the other three (marked by black arrows). When the outmost {111} facets are flattened, the embryos quickly grow up, turn into full dislocations, and then expand toward central region. Dislocations on different {111} slip planes react and are joined by pinning points, as marked in Figure 3(d). Typical dislocation structures in the following stage are shown in Figure 3(e) and (f). Slipping of the jogged dislocations dominates the plastic deformation. There are two events giving birth to dislocation jogs, i.e., one is the intersection of dislocations at different slip planes, and the other is the directly cross-slipping of dislocation loops nucleated from contact surface. During dislocation evolution, extended dislocation segments shrink or expand, and change the dislocation morphology intensively. In bulk materials, dislocation jogs and junctions decrease the mobility of dislocations, whereas in nanoparticles they can only exist for a short time before terminating at surface, similar to the case in metallic thin film [31].

In addition to the surface, single-arm source is another important dislocation source in confined volume structures [31]. Figure 4 depicts typical single-arm dislocation sources inside the deformed nanoparticle. In Figure 4(a) ~ (c), two spiraling arms of a source are connected by a short {001}<110>

Lomer dislocation, and their other ends terminate at free surface. The dislocation arms glide on two parallel {111} slip planes, while the Lomer dislocation glides along {001} plane. In Figure 4(d) ~ (f), the Lomer dislocation of another source has larger line length, and two arms are jogged extended dislocations. During compression, the dislocation arms cross slip and revolve around the Lomer dislocations. With the evolution of deformation, both the extended dislocation and Lomer dislocation finally annihilate at surface.

In the nanoindentation of bulk copper, initial dislocation homogeneously nucleated under indenter [4, 27]. While in the compression of nanoparticle, initial dislocations are always heterogeneously intrigued under surface steps. Besides the nucleation events, dislocation morphologies are also different in these two cases. In nanoparticle under [111] compression, jogged dislocations with short extended segments dominate the defects structures. While in [111] indentation of bulk material, extended dislocations are not distorted heavily [5], and prismatic loops usually emanate from contact zone [28].

## 3.2 [110] Compression

Under [110] compression as shown in Figure 5(a), the loading curve increases continuously up to the yielding point, following a power-law function rather than the linear relation of [111] compression. Figure 5(b) demonstrates the evolution of contact stress and contact area with respect to compression depth. Before yielding point, three local peaks appear on the contact stress curve, and meanwhile stepwise increase emerges on the contact area curve. In this case, three (110) atom layers are flattened before yielding. Each time when one atom layer is flattened, contact area increase suddenly, leading to the drop of contact stress. When multiple atom layers are involved in contact, the classical Hertzian model can be used to capture the elastic response [24]. In this model, the load $F$ as a function of compression depth $\delta$ is given as

$$F = \frac{4}{3} E^* R^{1/2} \delta^{3/2},$$

(4)

where $E^*$ is the reduced modulus of the nanoparticle, $R$ is the radius of nanoparticle. The fitted modulus in [110] direction is ~ 176.1 GPa. For nanoindentation of bulk copper, the modulus in [110] is ~ 163 GPa [5]. Nanoparticle has a larger modulus than that of bulk material, similar to [111] compression. Figure 5(b) illustrates the comparisons between the simulation data and the Hertzian prediction. It is seen that the overall elastic behaviors can be approximately characterized by the Hertzian model.

In the case of [110] compression, Figure 6(a) depicts two types of <111>{110} available slip systems. Figure 6(b) shows different views of the initial surface configuration under indenter. Initial dislocation firstly nucleates beneath surface steps, as shown in Figure 6(c). In this stage, slip system of type I in Figure 6(a) is activated. Initial dislocations nucleate and glide in the {111} slip planes parallel to the compressive direction. Following the nucleation and propagation of leading partial dislocations, trailing partial dislocations also nucleate from contact surface, forming extended dislocation ribbons. These extended dislocation segments on four adjacent slip planes connect with each other, and compose a prismatic loop, which moves along its glide prism and then emanates from contact region, as shown in Figure 6(d) ~ (h). The prismatic loop is highlighted in Figure 6(i), and the intersection between neighboring extended ribbons is stair-rod dislocation.

After releasing the prismatic loop, dislocations are prone to nucleate on slip system II, shown in Figure 6(a). Extended dislocation ribbons nucleate from surface, glide across the center region, and finally exhaust at the lateral surface. After being impacted by these extended dislocations, the prismatic loop in center region collapses (Figure 7a), and finally is expelled out of the nanoparticle, as shown in Figure 7(b) and (c). In further compression, only the nucleation and gliding of extended dislocations repeat to sustain plastic deformation, and Figure 7(d) ~ (i) depict this atomic process. Surface ledges are formed at the position where dislocation exhausts, as marked in Figure 7(i). The annihilation of dislocations leaves nanoparticle starved of dislocations, which is different from indenting bulk materials in [110] orientation [5, 26].

### 3.3 [112] Compression

Figure 8(a) illustrates the loading response under [112] compression. Similar to [110] compression, load increase in elastic stage follows a power-law function. However, the transition from elastic stage to plastic stage is more smooth, and no evident load-drop can be observed. Figure 8(b) gives the variations of averaged contact stress and contact area. When compression depth is smaller than ~ 4.7 Å, contact stress fluctuates and the curve exhibits several apparent peaks. Over the rest of compression, the magnitude of stress fluctuation decreases, and the stress keeps around ~ 11.0 GPa. Owing to the small crystal plane spacing in this orientation, contact area is raised stepwise for compression depth smaller than ~ 5.0 Å. Contact area keeps increasing until compression depth approaches ~ 9.5 Å, then it maintains almost at a constant value. Since multiple (112) atom layers are involved in elastic stage, Hertzian contact model is utilized to fit the elastic regime, as shown in Figure 8(a). The fitted modulus is ~ 155.9 GPa.

For [112] compression, Figure 9 depicts the available slip system and the initial surface configuration. Since (112) planes are not close-packed, the initial (112) surface facet is rough and composed of [$\bar{1}10$] surface steps. As the compression progresses, initial dislocations nucleate beneath these surface steps, and grow on the vertical ($\bar{1}\bar{1}1$) slip planes. Four dislocation embryos are observed, as shown in Figure 9(c). Since there is no resolved shear stress on the vertical {111}-type slip planes, these dislocation embryos are suppressed to expand. It is noted that nucleation of these small embryonic defects does not induce a significant load drop. In subsequent deformation, new nucleated dislocations prefer to glide and expand on the inclined (111) slip planes, which intrigues the yielding of nanoparticle, as shown in Figure 9(d). Upon further compression, partials generate at the intersections between stacking fault and surface, then glide on adjacent slip plane, and change the existing stacking fault to extrinsic fault. The migration of subsequent dislocations gradually broadens these nano-twinned structures, as shown in Figure 9(e) and (f). After the twin boundary (TB) departs from the contact region, perfect partial dislocation loops directly nucleate on the twin plane, expand and annihilate at surface, which is depicted in Figure 9(g) ~ (i). This results in the migration of TBs toward center. When two TBs from the top and bottom regions get close to each other, dislocations begin to nucleate at contact surface, and glide on slip planes inclined to TBs, as show in Figure 9(j) ~ (l). The existing TBs serve as barriers to these new dislocations. A few dislocations may slip across the TBs, and make TBs defective [34]. Under [112] compression, owing to the orientation of the activated <110>{111} slip system, migration of TBs mainly sustains the plastic deformation.

## 4. Conclusions

We perform MD simulations to investigate the compression of fcc copper nanoparticles in different directions. The results demonstrate that both the elastic and plastic behavior vary with compression orientation. Discrete atomic-scale surface steps play crucial roles in the mechanical response of nanoparticles. Under [111] compression, the elastic behavior follows the flat punch contact model. Under [110] and [112] compression, the classical Hertzian contact model is applicable in elastic regime. In nanoparticles, heterogeneous dislocations nucleate around surface steps and initiate the yield, different from the homogeneous dislocations under indenter tip for bulk materials. In severe compression, jogged dislocations dominate the plastic deformation in nanoparticles under [111] compression. For [110] compression, the plasticity is governed by extended dislocation ribbons. The annihilation of dislocations at surface leaves nanoparticle starved of dislocations. In [112] compression, deformation twinning is the major plastic deformation mechanism. This work reveals new atomic-scale features of mechanical response of metallic fcc nanoparticles.

## Acknowledgments

Supports from the National Natural Science Foundation of China (Grant Nos. 11525209) are acknowledged.

**Figures:**

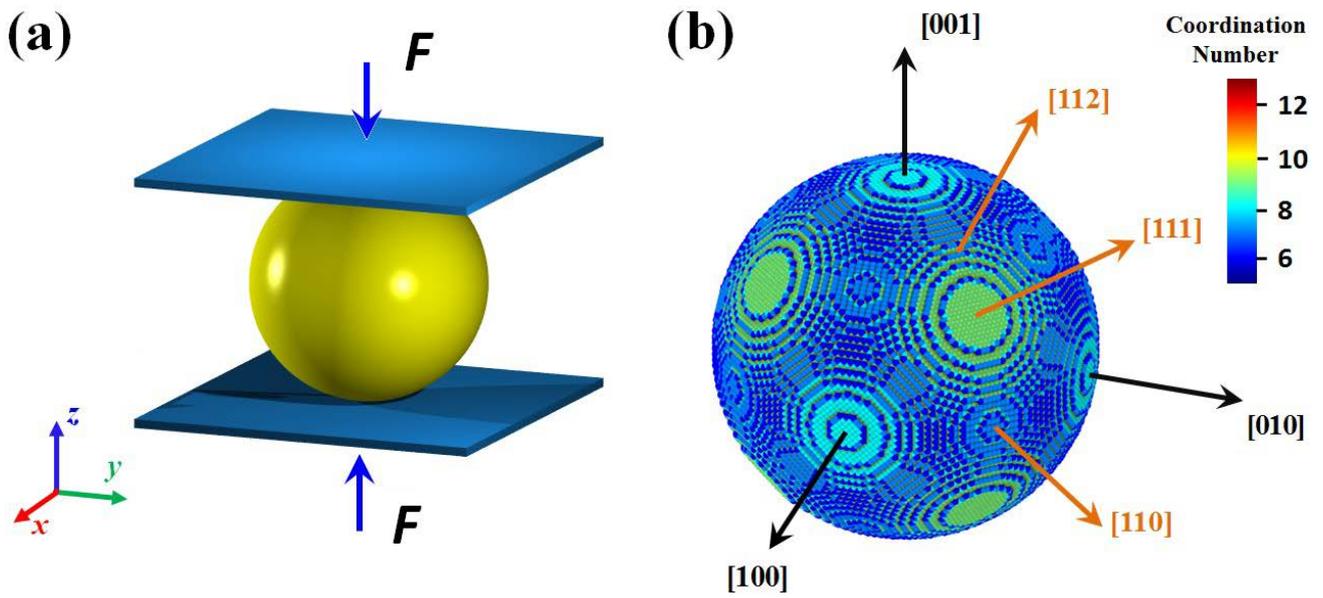

**Figure 1** (a) Schematic of the uniaxial compression of nanoparticle, and (b) initial configuration of a copper nanoparticle (surface atoms are colored by their coordination number)

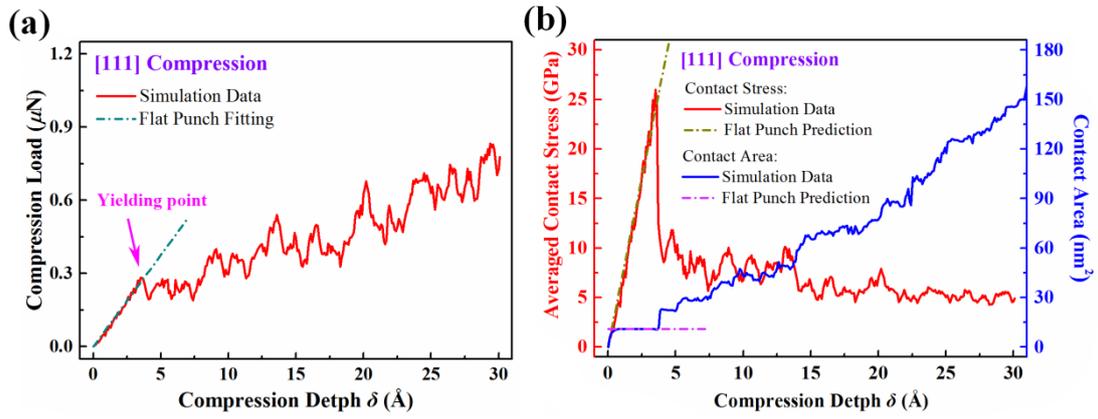

**Figure 2** (a) load response under [111] compression, and (b) variation of averaged contact stress and contact area with respect to compression depth

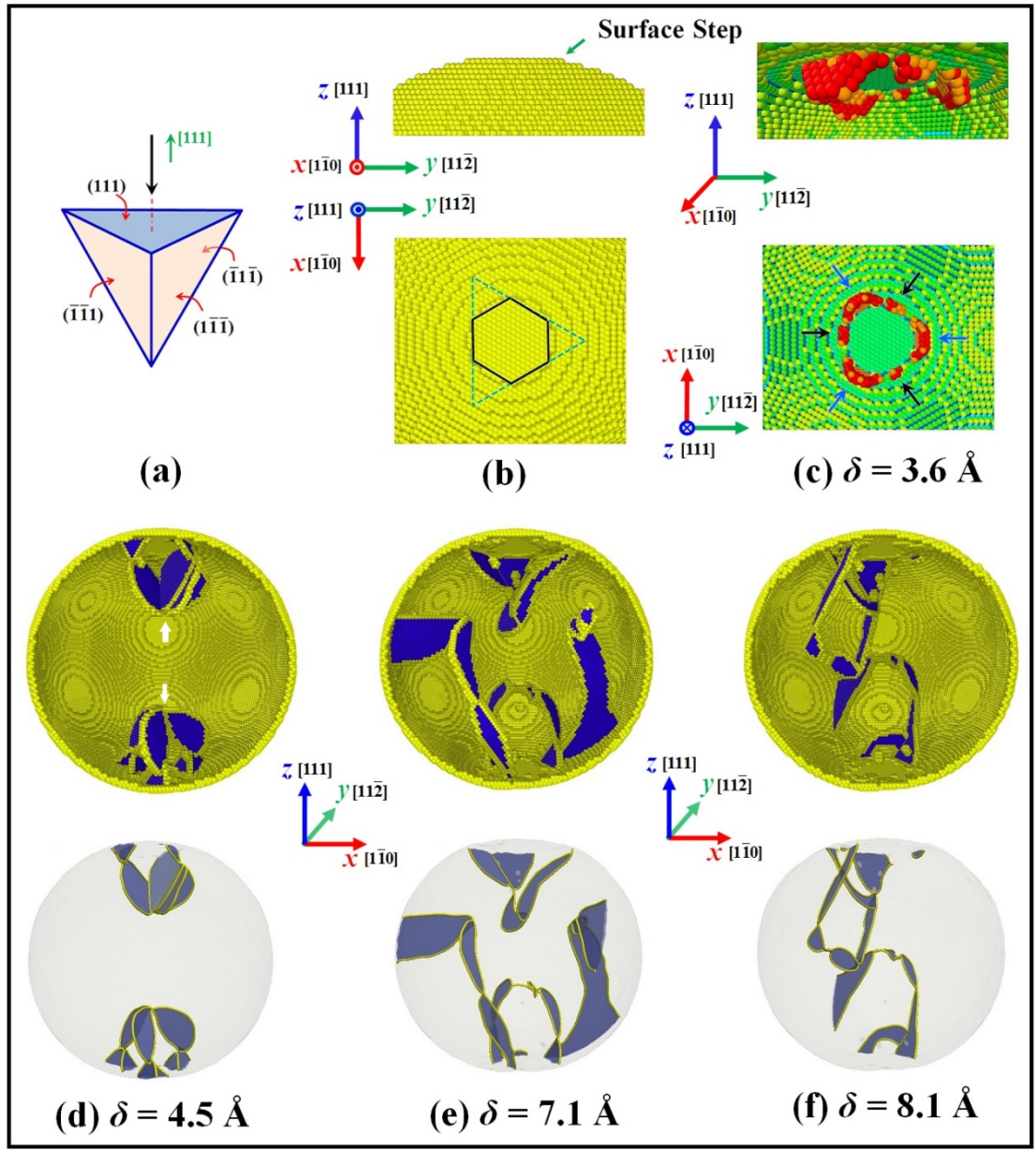

**Figure 3** Atomic structure and defect evolution under [111] compression. (In panel (c), atoms are colored by coordination number, and those with 12 are not shown for clarify. In panel (d) ~ (f), top parts are cross section views, atoms are colored by CNA parameter, and those in preface lattice are not shown. Atoms in yellow represent surface and dislocation cores, atoms in blue are in hcp lattice. Lower parts are the whole views of nanoparticle with only dislocation, stacking fault and surface are extracted and shown)

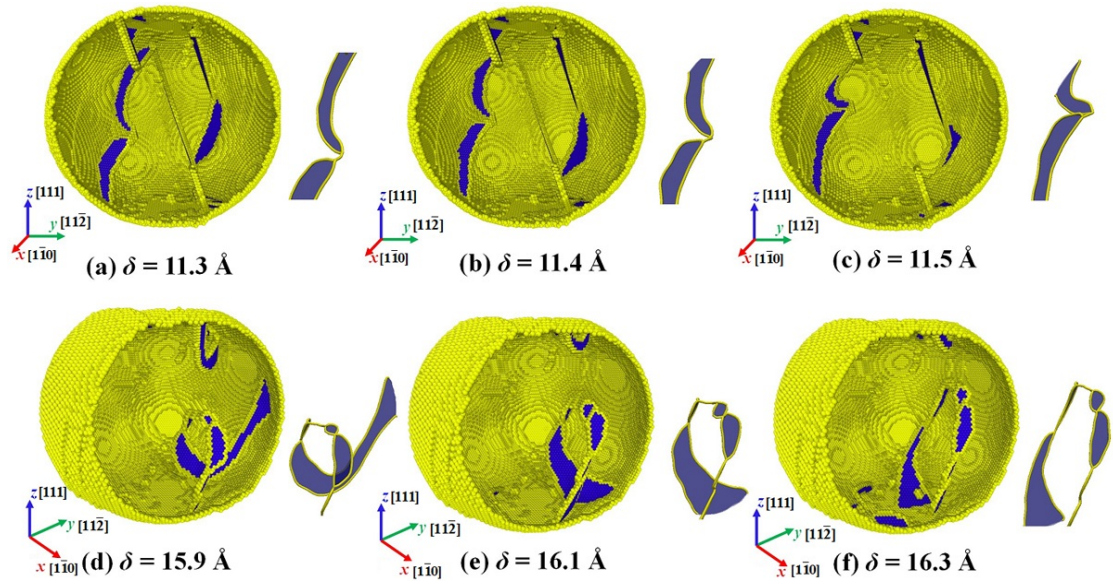

**Figure 4** Single-arm dislocation source inside nanoparticle under [111] compression (Atoms are colored by CNA parameter, and color scheme is the same as Figure 3)

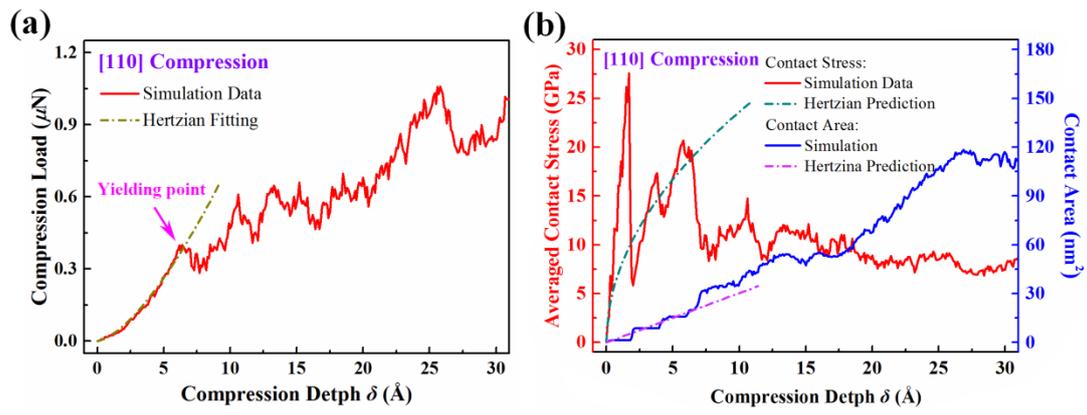

**Figure 5** (a) load response under [110] compression, and (b) variation of averaged contact stress and contact area with respect to compression depth

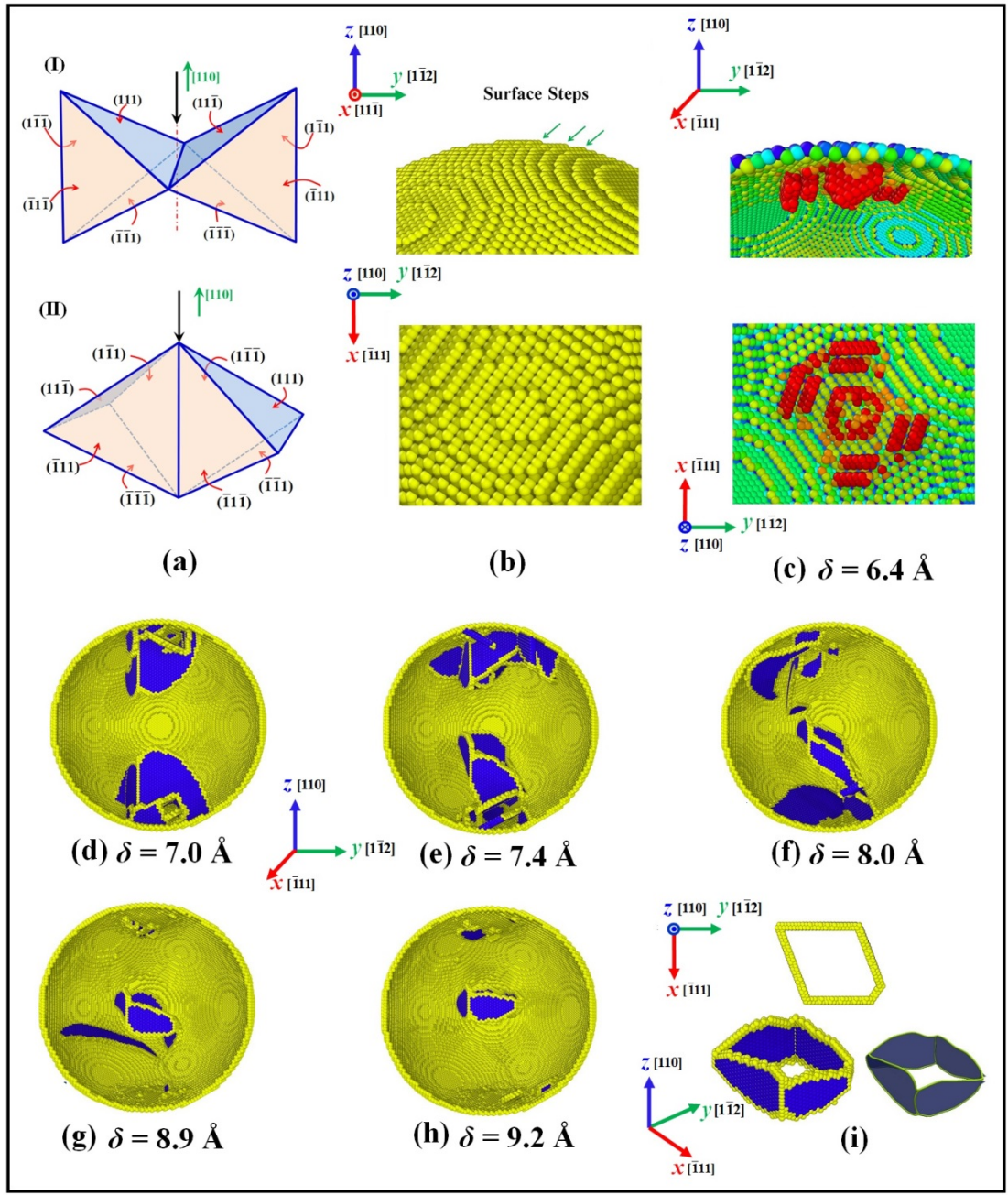

**Figure 6** Atomic structures and evolution of prismatic dislocation loop in nanoparticles under [110] compression (In panel (c), atoms are colored by coordination number. In panel (d) ~ (h), atoms are colored by CNA. Color scheme is the same as in Figure 3)

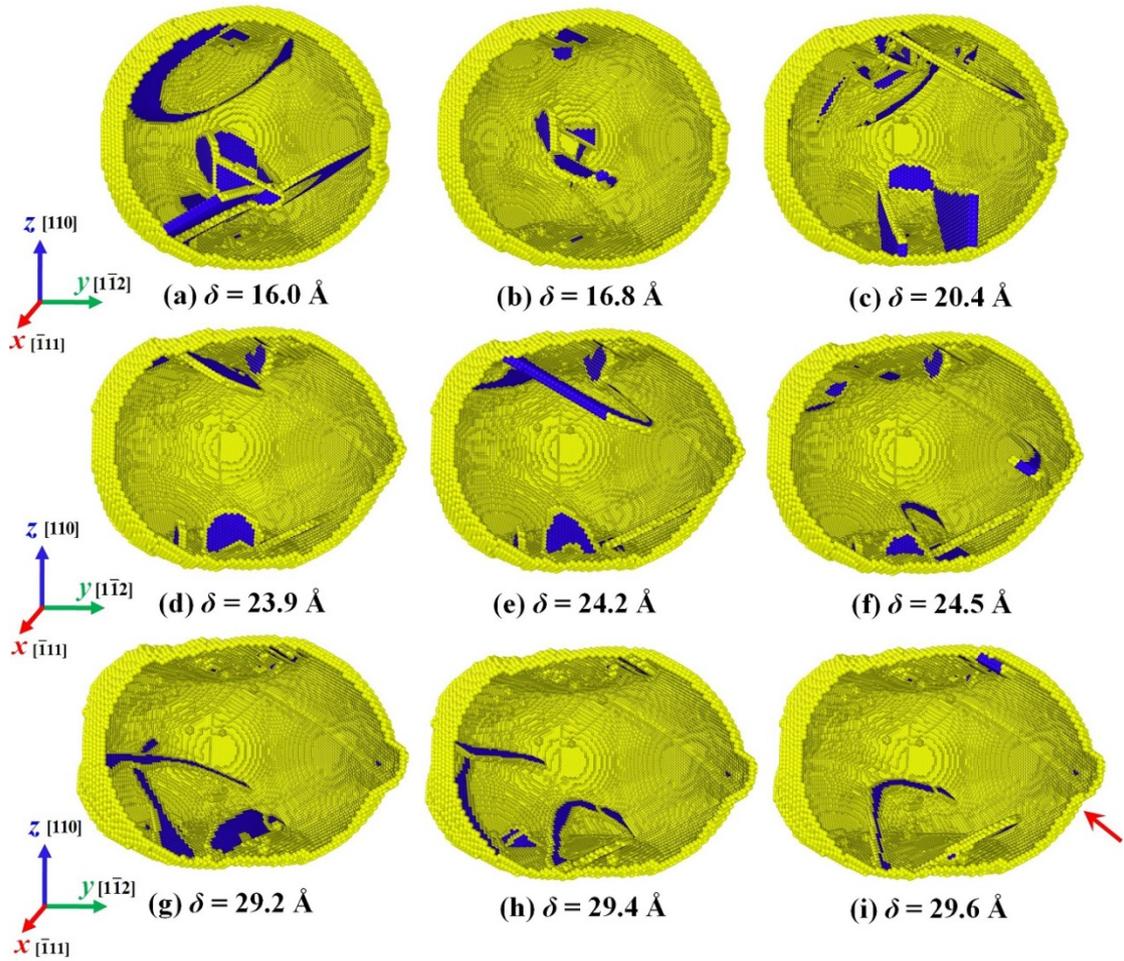

**Figure 7** Nucleation and exhaustion of extended dislocation ribbons in nanoparticle under [110] compression (Atoms are colored by CNA parameter, and color scheme is the same as Figure 3)

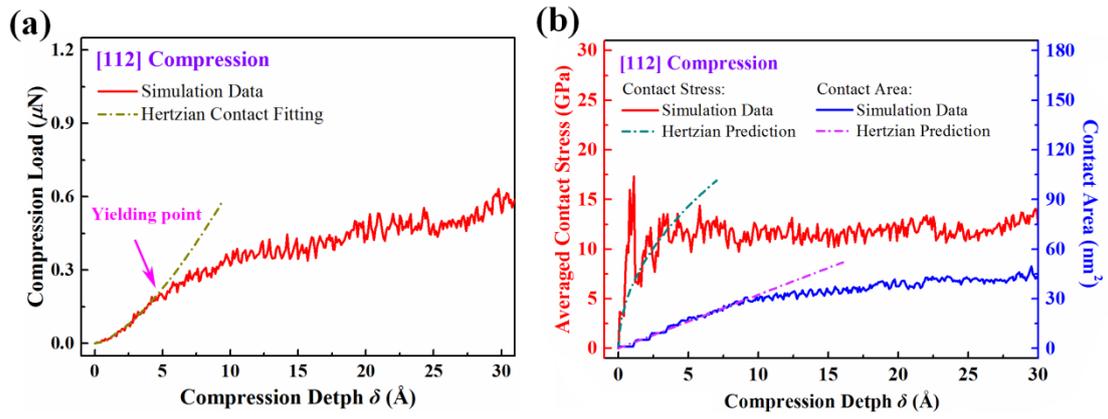

**Figure 8** (a) load response under [112] compression, and (b) variation of averaged contact stress and contact area with respect to compression depth

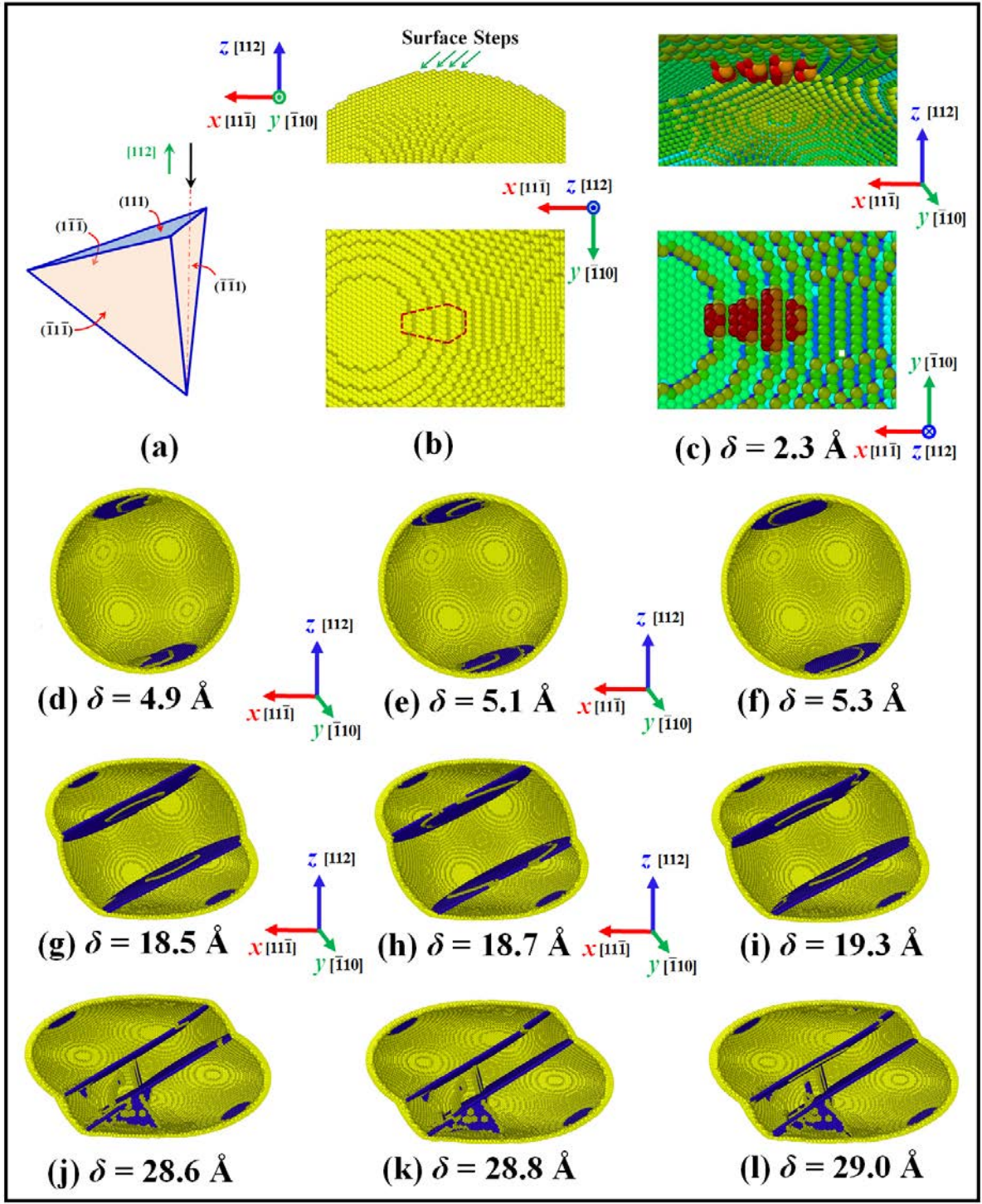

**Figure 9** Atomic structures and twin boundary migration in nanoparticle under [112] compression (In panel (c), atoms are colored by coordination number. In panel (d) ~ (l), atoms are colored by CNA parameters. Color scheme is the same as in Figure 3)